\input{aipcheck}
\documentclass[
    ,final          
]{aipproc}
\usepackage{epsfig}

\layoutstyle{8x11double}

\setcitestyle{numbers}

\newcommand{\eq}[1]{Eq.~(\ref{#1})}

\newcommand{\be}{\begin{equation}}
\newcommand{\ee}{\end{equation}}
\newcommand{\bea}{\begin{eqnarray}}
\newcommand{\eea}{\end{eqnarray}}
\newcommand{\ben}{\begin{eqnarray*}}
\newcommand{\een}{\end{eqnarray*}}

\newcommand{\DS}{Dyson-Schwinger }
\newcommand{\BS}{Bethe-Salpeter }
\newcommand{\ST}{Slavnov-Taylor }
\newcommand{\YM}{Yang-Mills }

\newcommand{\w}{\omega}
\newcommand{\e}{\varepsilon}
\newcommand{\al}{\alpha}
\newcommand{\ba}{\beta}
\newcommand{\ga}{\gamma}
\newcommand{\G}{\Gamma}
\newcommand{\de}{\delta}

\newcommand{\si}{\sigma}

\newcommand{\cs}{{\cal S}}
\renewcommand{\div}{\vec{\nabla}}

\newcommand{\s}[2]{{#1}\!\cdot\!{#2}}
\newcommand{\ov}[1]{\overline{#1}}

\begin{document}

\title{Heavy quarks, gluons and the confinement potential in 
Coulomb gauge}


\author{Carina Popovici}{
  address={Institut f\"ur Theoretische Physik, Universit\"at
T\"ubingen, Auf der Morgenstelle 14, D-72076 T\"ubingen, Germany}
}
\author{Peter Watson}{
  address={Institut f\"ur Theoretische Physik, Universit\"at
T\"ubingen, Auf der Morgenstelle 14, D-72076 T\"ubingen, Germany}
}
\author{Hugo Reinhardt}{
  address={Institut f\"ur Theoretische Physik, Universit\"at
T\"ubingen, Auf der Morgenstelle 14, D-72076 T\"ubingen, Germany}
}

\begin{abstract}
We consider the heavy quark limit of Coulomb gauge QCD, with the
truncation of the Yang-Mills sector to include only (dressed) 
two-point functions.  We find that the rainbow-ladder 
approximation to the gap and Bethe-Salpeter equations is 
nonperturbatively exact and moreover, we provide a direct 
connection between the temporal gluon propagator and the quark 
confinement potential. Further, we show that only  bound states of
color singlet quark-antiquark (meson) and quark-quark (SU(2) 
baryon) pairs are physically allowed.
\end{abstract}

\maketitle

Coulomb gauge is an ideal choice for investigating the confinement 
phenomenon. In this gauge, there is an appealing picture of 
confinement: the Gribov-Zwanziger scenario \cite{Gribov:1977wm}, 
whereby the temporal component of the gluon propagator provides a 
long-range confining force whereas the transverse spatial 
components are infrared suppressed (and therefore do not appear as 
asymptotic states). Confinement implies the existence of a 
nonperturbative scale (the string tension), and this naturally 
leads to the question: is there a simple connection between the 
(gauge fixed) Green's functions of \YM theory and this physical 
scale? To answer this question, we propose to study here the full 
(nonperturbative) QCD \BS equation in Coulomb gauge within a heavy 
quark mass expansion at leading order. In this truncated system, 
we will use the results inspired by the  lattice for the explicit 
Green's functions of the Yang--Mills sector  and in addition we 
will employ the \ST identity for the quark-gluon vertex.

We begin by considering the explicit quark contribution to the 
full QCD generating functional (unless otherwise specified, the 
Dirac and color indices in the fundamental representation are
implicit and we follow the conventions from 
\cite{Popovici:2010mb}):
\bea
&&Z[\ov{\chi},\chi]=\int{\cal D}\Phi\nonumber\\
&&\times\exp \left\{\imath
\int d^4x \ov{q}(x)\left[\imath\ga^0D_0
+\imath\s{\vec{\ga}}{\vec{D}}-m\right]q(x)
\right.
\nonumber\\&&
\left.
+\imath\int d^4x\left[\ov{\chi}(x) q(x)+
\ov{q}(x)\chi(x)\right]+\imath{\cal S}_{YM}\right\}.
\label{eq:genfunc}
\eea
$\cs_{YM}$ represents the \YM part of the action and the temporal 
and spatial components of the covariant derivative (in the 
fundamental color representation) are given by
\bea
D_0=\partial_{0}-\imath gT^a\si^a(x),\,\,\,\,\,\,
\vec{D}=\div+\imath gT^a\vec{A}^a(x),
\eea
where $\vec{A}$ and $\si$ refer to the spatial and temporal
components of the gluon field, respectively.

The full quark field $q$ ($\bar q$ is the conjugate field, 
and  $\chi$, $\ov{\chi}$ are the corresponding sources) is 
decomposed according to the heavy quark transformation:
\bea
&&\hspace{-0.7cm} q(x)=e^{-\imath mx_0}\left[h(x)+H(x)\right], 
\nonumber\\
&&\hspace{-0.7cm}h(x)=e^{\imath mx_0}\frac{{\bf 1}+\ga^0}{2}q(x),
\,\,\,\,\,\,
H(x)=e^{\imath mx_0} \frac{{\bf 1}-\ga^0}{2}q(x)
\label{eq:qdecomp}
\eea
(similarly for the antiquark field). We now insert this  
decomposition into the generating functional \eq{eq:genfunc}, 
integrate out the $H$-fields, and make an expansion in the heavy 
quark mass, similar to Heavy Quark Effective Theory [HQET] 
\cite{Neubert:1993mb}. At leading order, we get the following 
expression:
\bea 
&&\!\!\!\! \!\!\!\!\!\!\!\! \!\!\! \! Z[\ov{\chi},\chi]
=\int{\cal D}\Phi 
\exp
 \left\{\imath
\int d^4x\ov{h}(x)
\left[\imath\partial_{0x}+gT^a\si^a\right]
h(x)\right. \nonumber\\&&\!\!\!\! \!\!\!\!\!\!\!\!\!\!\!\!
+\left.\imath
\int d^4x\left[e^{-\imath mx_0}\ov{\chi}(x) h(x)
+e^{\imath mx_0}\ov{h}(x)\chi(x)\right]
+\imath{\cal S}_{YM}\right\}\nonumber\\&&\!\!\!\! \!\!\!\!\!\!\!\! 
\!\!\!\!
+{\cal O}\left(1/m\right).
\label{eq:genfunc4}
\eea
In the above, we have retained the full quark and antiquark 
sources (unlike HQET). This means that we can use the full gap and
\BS equations of QCD replacing, however, the kernels, propagators 
and vertices by their leading order expression in the mass 
expansion. Also, notice that the spin degrees of freedom have 
decoupled from the system.

In full Coulomb gauge QCD, the quark gap equation  is given by 
\footnote{This expression is in the second order formalism and can 
be directly inferred from the result obtained within the first 
order formalism \cite {Popovici:2008ty}.}:
\bea
&&\G_{\ov{q}q}(k)=\G_{\ov{q}q}^{(0)}(k)+\frac{1}{(2\pi)^4}
\int d^4 \w\nonumber\\&&
\times\left\{\G_{\ov{q}q\si}^{(0)a}W_{\ov{q}q}
(\w)\G_{\ov{q}q\si}^{b}(\w,-k,k-\w)W_{\si\si}^{ab}(k-\w)
\right.\nonumber\\&&\left.
+\G_{\ov{q}qA i}^{(0)a}W_{\ov{q}q}(\w)
\G_{\ov{q}qA j}^{b}(\w,-k,k-\w)W_{AAij}^{ab}(k-\w)
\right\}\nonumber\\
\label{eq:gap}
\eea
($\G$'s denote the various proper functions, $W$ denotes 
propagators, see \cite{Popovici:2008ty}). The gap equation is 
supplemented by the Coulomb gauge \ST identity 
\cite{Popovici:2010mb}:
\bea
&&k_3^0\G_{\ov{q}q\si}^{d}(k_1,k_2,k_3)=
\imath\frac{k_{3i}}
{\vec{k}_3^2}\G_{\ov{q}qA i}^{a}(k_1,k_2,k_3)
\G_{\ov{c}c}^{ad}(-k_3)\nonumber\\&&
+\G_{\ov{q}q}(k_1)\left[\tilde{\G}_{\ov{q};\ov{c}cq}^{d}
(k_1+q_0,k_3-q_0;k_2)+\imath gT^d\right]
\nonumber\\&&
+\left[\tilde{\G}_{q;\ov{c}c\ov{q}}^{d}(k_2+q_0,k_3-q_0;k_1)-
\imath gT^d\right]\G_{\ov{q}q}(-k_2) 
\label{eq:stid}
\eea
where $k_1+k_2+k_3=0$, $q_0$ is an arbitrary energy injection 
scale (arising from the noncovariance of Coulomb gauge),
$\G_{\ov{c}c}$ is the ghost proper two-point function,
$\tilde{\G}_{\ov{q};\ov{c}cq}$ and $\tilde{\G}_{q;\ov{c}c\ov{q}}$ are
ghost-quark kernels associated with the Gauss-BRST transform 
(see also \cite{Watson:2008fb}). 

Now, as a consequence of the (Coulomb gauge) decomposition,
\eq{eq:qdecomp} and under the assumption that the pure \YM 
vertices may be neglected (retaining, however, the dressed gluon 
propagator), the \DS equation for the nonperturbative spatial 
quark-gluon vertex  furnishes the result that 
$\G_{\bar qqA}\sim O(1/m)$ (see \cite{Popovici:2010mb} for a
complete discussion and justification of this truncation). 
Similarly, the ghost-quark kernels can be neglected. Thus, under 
our truncation scheme, the \ST identity reduces to
\bea
&&k_3^0\G_{\ov{q}q\si}^{d}(k_1,k_2,k_3)=\nonumber\\&&
\imath g\left[ \G_{\ov{q}q}(k_1) T^d-T^d\G_{\ov{q}q}(-k_2)\right]
+{\cal O}\left(1/m\right).
\eea
This is then inserted into \eq{eq:gap}, together with the 
tree-level quark  proper two-point function
\be
\G_{\ov{q}q}^{(0)}(k)=\imath\left[k_0-m\right]
+{\cal O}\left(1/m\right)
\label{eq:gaptree}
\ee
and the tree level quark gluon vertex 
\be
\G_{\ov{q}q\si}^{(0)a}(k_1,k_2,k_3)= gT^a+{\cal O}\left(1/m\right)
\label{eq:feyn0}
\ee
that follow from the generating functional \eq{eq:genfunc4}. The 
general form of the nonperturbative temporal gluon propagator is
given by \cite{Watson:2007vc}:
\be
W_{\si\si}^{ab}(\vec k)=
\de^{ab}\frac{\imath}{\vec{k}^2}D_{\si\si}(\vec{k}^2).
\label{eq:Wsisi}
\ee
Lattice results \cite{Quandt:2008zj} motivate that the dressing 
function $D_{\si\si}$ is largely independent of energy and likely 
to behave as $1/\vec{k}^2$ for vanishing $\vec{k}^2$ (the explicit 
form of $D_{\si\si}$ will only be needed in the last step of the 
calculation). Putting  all this together, we find the following
solution to \eq{eq:gap} for the heavy quark propagator:
\bea
W_{\ov{q}q}(k_0)=\frac{-\imath}{\left[k_0-m-
{\cal I}_r+\imath\e\right]}+{\cal O}\left(1/m\right),
\label{eq:quarkpropnonpert}
\eea
where the (implicitly regularized, denoted by ``$r$'') constant is
given by
\be
 {\cal I}_r =\frac{1}{2 (2\pi)^3}g^2C_F
\int_r\frac{d\vec{\w}D_{\si\si}(\vec{\w})}{\vec{\w}^2}
+{\cal O}\left(1/m\right),
\label{eq:iregularized}
\ee
with  the Casimir factor $C_F=(N_c^2-1)/2N_c$. When solving 
\eq{eq:gap}, the ordering of the integration is set such that the 
temporal integral is performed first, under the condition that the
spatial integral is regularized and finite. Inserting the solution
\eq{eq:quarkpropnonpert} into the \ST identity, we find that the 
temporal quark-gluon vertex remains nonperturbatively bare:
\be
\G_{\ov{q}q\si}^{a}(k_1,k_2,k_3)=gT^a+{\cal O}\left(1/m\right).
\label{eq:qqsinp}
\ee

Note that the propagator \eq{eq:quarkpropnonpert} has a single pole
in the complex $k_0$-plane (due to the mass expansion) and 
therefore it is necessary to explicitly define the Feynman 
prescription. From \eq{eq:quarkpropnonpert} it then follows that 
the closed quark loops (virtual quark-antiquark pairs) vanish due 
to the energy integration, which implies that the theory is 
quenched in the heavy mass limit:
\be
\int\frac{dk_0}{\left[k_0-m- {\cal I}_r+\imath\e\right]
\left[k_0+p_0-m- {\cal I}_r+\imath\e\right]}=0.
\label{eq:tempint}
\ee
We also emphasize that the position of the  pole has no physical 
meaning since the quark can never be on-shell. The poles in the 
quark propagator are situated at infinity (thanks  to ${\cal I}_r$ 
as the regularization is removed) meaning that either one requires
infinite energy to create a quark from the vacuum or, if a hadronic
system is considered, only the relative energy (derived from the 
\BS equation) is important.

The solution for the antiquark propagator reads:
\bea
W_{q\ov{q}}(k_0)=\frac{-\imath}{\left[k_0+m-
{\cal I}_r+\imath\e\right]}+{\cal O}\left(1/m\right).
\label{eq:antiquarkpropnonpert}
\eea
Notice the assignment of the Feynman prescription, similar to
\eq{eq:quarkpropnonpert} -- see also the discussion in 
\cite{Popovici:2010mb}. This has the important consequence that 
the \BS equation for the quark-antiquark states has a physical 
interpretation of a bound state equation (see below). 
The corresponding vertex reads:
\be
\G_{q\ov{q}\si}^{a}(k_1,k_2,k_3)=-gT^a+{\cal O}\left(1/m\right).
\label{eq:antiqqsinp}
\ee

Let us now consider the full homogeneous \BS equation for 
quark-antiquark bound states:
\bea
\G(p;P)_{\al\ba}&=&-\frac{1}{(2\pi)^4}\int d kK_{\al\ba;\de\ga}
(p,k;P)\nonumber\\
&\times&\left[W_{\ov{q}
q}(k_+)\G(k;P)W_{\ov{q}q}(k_-)\right]_{\ga\de}
\label{eq:bseq}
\eea
where $k_\pm$ (similarly for $p_\pm$) are the momenta of the quarks
(with the notations from \cite{Popovici:2010mb}), $P$ is the  
4-momentum of the bound state (assuming that a solution exists),
$K$ represents the \BS kernel and $\G$ is the \BS vertex function
for the particular bound state under consideration. Further, we 
explicitly identify the antiquark contribution, i.e. 
$W_{\ov{q}q}(k_-)=-W_{q\ov{q}}^T(-k_-)$ (also in the kernel).

When constructing the \BS kernel, we use the fact that the 
temporal integration performed over multiple quark propagators with
the same relative sign for the Feynman prescription vanishes 
(similar to \eq{eq:tempint}, but in this case the terms originate 
from internal quark or antiquark propagators) and hence, the kernel
reduces to the ladder truncation \cite{Popovici:2010mb}:
\be
K_{\al\ba;\de\ga}(p,k)= 
\G_{\bar qq\si\al\ga}^{a}W_{\si\si}^{ab}(\vec k) 
\G_{q\bar q\si\ba\de}^{Tb}.
\label{eq:kernel}
\ee

We now insert the nonperturbative results for the propagators and 
vertices, Eqs.~(\ref{eq:quarkpropnonpert},\ref{eq:qqsinp},
\ref{eq:antiquarkpropnonpert},\ref{eq:antiqqsinp})
and take the form, \eq{eq:Wsisi}, for the temporal gluon 
propagator. Further, we notice that that the \BS equation is 
independent of the relative quark energy and hence we can perform 
the temporal integration over the quark propagators. The energy 
integration over the quark and antiquark propagators now leads to 
(unlike \eq{eq:tempint}):
\bea
\!\int_{-\infty}^\infty\!\!
\frac{dk_0}{\left[k_+^0-m-{\cal I}_r+\imath\e\right]\left[k_-^0-m+
{\cal I}_r-\imath\e\right]}\!=\!\frac{-\pi\imath}{{\cal I}_r}.
\eea
By using  the expression \eq{eq:iregularized} for ${\cal I}_r$, and
Fourier transforming to coordinate space, we find the following
simple energy solution for the pole condition of the \BS equation
\eq{eq:bseq}:
\be
P_0=\frac{g^2}{(2\pi)^3}\int_{r}\frac{d \vec{\w}D_{\si\si}
(\vec{\w})}{\vec{\w}^2}
\left\{C_F-e^{\imath\vec{\w}\cdot\vec{y}}C_M\right\}+{\cal O}
\left(1/m\right).
\label{eq:p0sol}
\ee
where $C_M$ arises from the color structure and is yet to be
identified ($\G$ is not assumed to be a color singlet):
\be
\left[T^a\G(\vec{y})T^a\right]_{\al\ba}=C_M\G_{\al\ba}(\vec{y}).
\ee 

Because the total color charge of the system is conserved and
vanishing \cite{Reinhardt:2008pr}, a single quark (or antiquark)
cannot be prepared in isolation. Thus, the bound state energy $P_0$
can only be either confining for large separations, i.e. increase
linearly with the separation between the quark and antiquark, or be
infinite when the hypothetical regularization is removed (so that 
the system  cannot be physically created). If the temporal gluon 
propagator dressing function is more infrared divergent than 
$1/|\vec{\w}|$,  then 
\be
C_F=C_M
\label{eq:cfcm}
\ee
is required such that the spatial integral is convergent. This 
gives the condition
\be
\G_{\al\ga}(\vec{y})=\de_{\al\ga}\G(\vec{y}),
\ee
which means that the quark-antiquark \BS equation can only have a
finite solution for \emph{color singlet} states and otherwise the 
energy of the system is divergent. Assuming that in the infrared
$D_{\si\si}=X/\vec{\w}^2$ (as indicated by the lattice) where $X$ 
is some combination of constants, then from \eq{eq:p0sol} with the
condition \eq{eq:cfcm} we find
\be
P_0\equiv\si|\vec{y}|=\frac{g^2C_FX}{8\pi}|\vec{y}|+{\cal O}
\left(1/m\right).
\ee
The above result is that there exists a direct connection between 
the string tension $\si$ and the nonperturbative Yang--Mills sector
of QCD at least under the truncation scheme considered here.  

A similar calculation performed for the diquark \BS equation shows
that the diquarks are confined for $N_c=2$ colors, corresponding to
the $SU(2)$ baryon, and otherwise there are no (finite) physical 
states.

\begin{theacknowledgments}
C.P. has been supported by the Deutscher Akademischer Austausch 
Dienst (DAAD). P.W. and H.R. have been supported by the Deutsche
Forschungsgemeinschaft (DFG) under contracts no. DFG-Re856/6-2,3.
C.P. thanks the organizers, in particular  F. Llanes-Estrada, for 
the support.
\end{theacknowledgments}

\end{document}